\documentclass[12pt]{article}
\usepackage{latexsym}

\usepackage[tbtags]{amsmath}
\usepackage{epsfig,amstext,amssymb,amsthm,latexsym}
\usepackage{amssymb,color}

\definecolor{c20}{rgb}{0.,0.7,0.}
\definecolor{c30}{rgb}{0.,0.,1.}
\definecolor{c40}{rgb}{1,0.1,0.7}
\definecolor{c50}{rgb}{1,0,0}

\date{}
\newtheorem{theorem}{Theorem}[section]
\newtheorem{lemma}{Lemma}[section]
\newtheorem{proposition}{Proposition}[section]

\newtheorem{remark}{Remark}[section]
\newtheorem{example}{Example}[section]

\newtheorem{definition}{Definition}[section]

\def\eqref#1{ (\ref{#1})}

\makeatletter 
\@addtoreset{equation}{section}
\makeatother 
 
\begin{document}

\title{On the Use of the Borel-Cantelli Lemma in Markov Chains}

\author{ Alexei Stepanov,\thanks{\noindent Immanuel Kant Baltic Federal University, A.Nevskogo 14, Kaliningrad, 236041 Russia; e-mail: {\it alexeistep45@mail.ru}}~
\small{\it Immanuel Kant Baltic Federal University, Russia}}

\def\abstractname{}

\date{\begin{abstract} In the present paper, we propose  technical generalizations of  the Borel-Cantelli lemma. These generalizations can be further used to derive strong limit results for Markov chains. In our work, we obtain some strong limit results.
\end{abstract}}

\maketitle  \vspace{3mm}\noindent


\section{Introduction}

Suppose $A_1,A_2,\cdots$ is a sequence of events on a common
probability space and that $A^c_i$ denotes the complement of event
$A_i$. The Borel-Cantelli lemma,  presented here as Lemma~\ref{lemma1.1},
is used  for producing strong limit theorems.

\begin{lemma}\label{lemma1.1}
\begin{enumerate}
\item If, for any sequence  $A_1,A_2,\cdots$ of events,
$\sum_{n=1}^\infty P(A_n)<\infty,$
then $P(A_n\ i.o.)=0$, where i.o. is an abbreviation for
"infinitively often``.
\item If $A_1,A_2,\cdots$ is a sequence of
independent events and if
$\sum_{n=1}^\infty P(A_n)=\infty$, then $P(A_n\ i.o.)=1$.
\end{enumerate}
\end{lemma}

The first part of the Borel-Cantelli lemma is generalized in Barndorff-Nielsen (1961) and Balakrishnan and Stepanov (2010). These results are presented below as Lemma~\ref{lemma1.2} and Lemma~\ref{lemma1.3}, respectively.

\begin{lemma}\label{lemma1.2}
Let $A_1,A_2,\ldots$ be a sequence of events such that
$P(A_n)\rightarrow 0$. If $\sum_{n=1}^\infty P(A_n A^c_{n+1})<\infty$,
then $P(A_n\ i.o.)=0$.
\end{lemma}

\begin{lemma}\label{lemma1.3}
Let  $A_1,A_2,\ldots$ be a sequence of events such that
$P(A_n)\rightarrow 0$.  If, for some $m\geq 0$,
$\sum_{n=1}^\infty P(A^c_n\ldots A_{n+m-1}^cA_{n+m})<\infty$,
then $P(A_n\ i.o.)=0$.
\end{lemma}

Many publications were devoted to the second part of the Borel-Cantelli lemma in attempts to
weaken the independence condition.  Erd\"os and R\'enyi (1959) discovered that this condition can be replaced by the weaker condition of pairwise independence of events
$A_1, A_2, \ldots$  They also found that the later condition, in its turn,  can be replaced by the  condition: $P(A_iA_j)\leq P(A_i)P(A_j)$ for every $i\not=j$.
Further generalizations of the Borel–Cantelli lemma were obtained independently by Kochen and Stone (1964) and
Spitzer (1964). Lamperti (1963) formulated the following proposition. If $C$ is a positive constant, $\sum_{n=1}^\infty P(A_n)=\infty$
and $P(A_iA_j)\leq CP(A_i)P(A_j)$ for all large enough $i\not=j$,
then $P(A_n\ i.o.)>0$. Petrov (2002) showed that if $C\geq 1$, then $P(A_n\ i.o.)\geq 1/C$. Recently Frolov (2012) extended the results of Petrov (2002), (2004) and found more sophisticated lower bounds for $P(A_n\ i.o.)$. The second part of the Borel-Cantelli lemma was also discussed in the works of Chang and Erd\"os (1952), Kounias (1968), M\'ori and Sz\'ekely (1983), Martikainen and Petrov (1990), and Petrov (1995). For a review on the Borel-Cantelli lemma, one may  refer to the book of Chandra (2012).

An alternative way to get around the independence assumption  in the second part of the Borel-Cantelli lemma is to use conditioning, as was proposed in L\'evy (1937).
\begin{lemma}\label{lemma1.4} Let $A_1, A_2, \ldots$ be a sequence of events and $I_{A_n}$  the indicator function of the event $A_n$. Then with probability one
$$
\sum_{n=1}^\infty I_{A_n}=\infty\quad \mbox{iff}\quad \sum_{n=2}^\infty P(A_{n}\mid \sigma\{A_1,\ldots,A_{n-1}\})=\infty.
$$ 
\end{lemma}
L\'evy's lemma implies the classical form of the Borel-Cantelli lemma.\\

In our work, we are going to formulate a version of the Borel-Cantelli lemma which is applicable to Markov chains. Let us define the sequences of events with Markov property. 
\begin{definition}\label{definition1.1}
We say that $A_n\ (n\geq 1)$ is a Markov sequence of events if the sequence of random variables $I_{A_n}\ (n\geq 1)$ is a Markov chain.
\end{definition}

Obviously, Markov sequences of events are associated with Markov chains.

The rest of our paper is organized as follows. In Section~2, we present  technical generalizations of  the Borel-Cantelli lemma. These generalizations are stated in terms of Markov sequences of events and can be further applied to the asymptotic theory of Markov chains. In Section~3, the results of Section~2 are used to derive  strong limit results for  Markov chains.

\section{The Borel-Cantelli Lemma for Markov Sequences of Events}

In this section, we consider Markov sequences of events and discuss  conditions for the validity of $P(A_n\ i.o.)=0/1$. It is easy to show that the condition $P(A_n)\not\rightarrow 0$  implies that $P(A_n\ i.o.)>0$. In the following generalization  of the Borel-Cantelli lemma we assume that $P(A_n)\rightarrow 0$.  

\begin{lemma}\label{lemma2.1}
Let  $A_1, A_2, \dots$ be a Markov sequence of events such that $P(A_n)\rightarrow 0$, and $N$ be a number such that $P(A_n)\not=1$ for $n\geq N$. Let us consider the series
\begin{equation}\label{2.1}
\sum_{n=N}^\infty P(A_{n+1}\mid A_n^c).
\end{equation}
If the series in (\ref{2.1}) is  convergent, then $P(A_n\ i.o.)=0$. If the series in (\ref{2.1}) is divergent, then $P(A_n\ i.o.)=1$.
\end{lemma}

\begin{gproof}{} Let us choose $n\geq N$. It follows from the Markov property that
$$
1-P(A_n\cup A_{n+1}\cup \ldots\cup A_{n+k})=P(A_n^cA_{n+1}^c\ldots A_{n+k}^c)=
$$$$
P(A_n^c)P(A_{n+1}^c\mid A_n^c)\ldots P(A_{n+k}^c\mid A_{n+k-1}^c).
$$
Then
\begin{equation}\label{2.2}
1-P(\cup_{i=n}^{\infty }A_i)=e^{\log P(A_n^c)+\sum_{i=n}^\infty \log P(A_{i+1}^c\mid A_i^c) }.
\end{equation}
Since $\log(1-x)\sim -x$ as $x\rightarrow0$, the convergence/divergence of the series $\sum_{i=n}^\infty  P(A_{i+1}\mid A_i^c)$ implies the convergence/divergence of the series $\sum_{i=n}^\infty \log P(A_{i+1}^c\mid A_i^c)$ (and vice verse). Observe that  $P(A_n\ i.o.)=\lim_{n\rightarrow \infty }P(\cup_{i=n}^{\infty }A_i)$. 
The result of Lemma~\ref{lemma2.1} readily follows  from (\ref{2.2}). 
\end{gproof}

\begin{remark}\label{remark2.1} It should be noted that the first statement in Lemma~\ref{lemma2.1} can be derived  from Lemma~\ref{lemma1.3} and the second statement in Lemma~\ref{lemma2.1} can be obtained from Lemma~\ref{lemma1.4}.
\end{remark}

Since $P(A_{n+1}\mid A_n^c)=\frac{P(A_n^cA_{n+1})}{P(A_n^c)}$ and in Lemma~\ref{2.1} $P(A_n^c)\rightarrow 1$, one can rewrite Lemma~\ref{lemma2.1} in the following way.

\begin{lemma}\label{lemma2.2}
 Let  $A_1, A_2, \dots$ be a Markov sequence of events such that $P(A_n)\rightarrow 0$. Let us consider the series
\begin{equation}\label{2.3}
\sum_{n=1}^\infty P(A_n^cA_{n+1}).
\end{equation}
If the series in (\ref{2.3}) is  convergent, then $P(A_n\ i.o.)=0$. If the series in (\ref{2.3}) is divergent, then $P(A_n\ i.o.)=1$.
\end{lemma}

Obviously, if the Markov property holds for some sequence $A_n$, then
$$
P(A_n^cA_{n+1}^c\ldots A_{n+k}^c)=
P(A_{n}^c\mid A_{n+1}^c)\ldots P(A_{n+k-1}^c\mid A_{n+k}^c)P(A_{n+k}^c).
$$
With little modifications in the above proofs, one can show the following result.
\begin{lemma}\label{lemma2.3}
 Let  $A_1, A_2, \dots$ be a Markov sequence of events such that $P(A_n)\rightarrow 0$. Let us consider the series
\begin{equation}\label{2.3}
\sum_{n=1}^\infty P(A_nA_{n+1}^c).
\end{equation}
If the series in (\ref{2.3}) is  convergent, then $P(A_n\ i.o.)=0$. If the series in (\ref{2.3}) is divergent, then $P(A_n\ i.o.)=1$.
\end{lemma}

We present some further simple generalizations of the Borel-Cantelli lemma which are applicable to the theory of Markov chains. It is known  that a sequence $X_1, X_2, \ldots$ forms a Markov chain of order $k$ (with memory $k$) if for all $n>k$
$$
P(X_n=x_n\mid X_{n-1}=x_{n-1},\ldots,X_1=x_1)=P(X_n=x_n\mid X_{n-1}=x_{n-1},\ldots,X_{n-k}=x_{n-k}).
$$
\begin{definition}\label{definition2.1}
We say that $A_n\ (n\geq 1)$ is a Markov sequence of events of order $k$ if the sequence of random variables $I_{A_n}\ (n\geq 1)$ is a Markov chain of order $k$.
\end{definition}
Lemma~\ref{lemma2.1} and Lemma~\ref{lemma2.2} can be easily extended to such sequences.

\begin{lemma}\label{lemma2.4}
Let  $A_1, A_2, \dots$ be a Markov sequence of events of order $k$ such that $P(A_n)\rightarrow 0$, and $N$ be a number such that $P(A_n)\not=1$ for $n\geq N$. Let us consider the series 
\begin{equation}\label{2.4}
\sum_{n=\max\{k,N\}}^\infty P(A_{n+1}\mid A_n^c\ldots A_{n-k+1}^c).
\end{equation}
If the series in (\ref{2.4}) is  convergent, then $P(A_n\ i.o.)=0$. If the series in (\ref{2.4}) is divergent, then $P(A_n\ i.o.)=1$.
\end{lemma}

\begin{lemma}\label{lemma2.5}
Let  $A_1, A_2, \dots$ be a Markov sequence of events of order $k$ such that $P(A_n)\rightarrow 0$. Let us consider the series 
\begin{equation}\label{2.5}
\sum_{n=1}^\infty P(A_n^c\ldots A_{n+k-1}^cA_{n+k}).
\end{equation}
If the series in (\ref{2.5}) is  convergent, then $P(A_n\ i.o.)=0$. If the series in (\ref{2.5}) is divergent, then $P(A_n\ i.o.)=1$.
\end{lemma}

\section{Applications}
In this section, we discuss two applications to the  results of Section~2. In Subsection~3.1, we derive a  strong limit theorem for the concomitants of maxima. In Subsection~3.2, we obtain  strong limit results for  the $F^\alpha$-scheme.

\subsection{Strong Limit Results for the Concomitants of Maxima}

Assume in this subsection that $(X,Y), (X_1,Y_1), (X_2,Y_2),\ldots,(X_n,Y_n)$
are independent  and identically distributed random vectors with
continuous bivariate distribution function $F(x,y)$ and  corresponding
marginal distributions functions $H(x)$ and $G(y)$.  In the case of existence the bivariate density of $(X,Y)$ will be denoted as $f(x,y)$. Let $X_{1,n}\leq
X_{2,n}\leq\ldots\leq X_{n,n}$ be the order statistics obtained
from the sample $X_1,X_2,\ldots,X_n$ and
$Y_{[1,n]},Y_{[2,n]},\ldots,Y_{[n,n]}$ the corresponding
concomitants of these order statistics, which relate to the sample
$Y_1,Y_2,\ldots,Y_n$. We will also  use the designations $Z_{[1]}=(X_{1,1},Y_{[1,1]}),\ldots,Z_{[n]}=(X_{n,n},Y_{[n,n]})$. The concept of concomitants of order
statistics is introduced in David (1973) and Bhattacharya (1974).

The limit behavior of $P(Y_{[n-k,n]}\leq y)$ when
$k\geq0$ is fixed and  $n\rightarrow\infty$ is studied in many
research works, see for example, David and
Galambos (1974),  (1987), Egorov and Nevzorov
(1984), David (1994), Goel and Hall (1994), and David and Nagaraja
(2003); see also the references therein.

Strong limit results for the concomitants of order statistics are rarely discussed. One can mention the publication of Goel and Hall (1994), where some strong limit theorems for differences  between order statistics and  concomitants are obtained, and the work of Sen (1981), in which the strong invariance principle for concomitants is discussed. In this subsection, to show the power of Lemma~\ref{lemma2.2}, we establish  a new strong limit theorem for
$Y_{[n,n]}$. This result can be further extended to the concomitants of top order statistics.

Let  $l_{H}=\inf\{x\in \mathbb{R}:\:H(x)>0\}$ and $r_{H}=\sup\{x\in \mathbb{R}:\:H(x)<1\}$ be the  left and right
extremities of $H$, respectively.

In the absolutely continuous case, when the bivariate density $f(x,y)$ exists, one can write the joint density of $Z_{[1]},\ldots,Z_{[n]}$ as
$$
f_{Z_{[1]},\ldots,Z_{[n]}}(x_1,y_1,\ldots,x_n,y_n)=f(x_1,y_n)\ldots f(x_n,y_n)\quad (x_1\leq \ldots\leq x_n,\ y_i\in \mathbb{R}).
$$
It follows that
$$
f_{Z_{[n+1]}\mid Z_{[n]}\ldots Z_{[1]}}(x_{n+1},y_{n+1}\mid x_n,y_n,\ldots ,x_1,y_1)=
f(x_{n+1},y_{n+1})\quad (x_{n+1}\geq x_n,\ y_n,y_{n+1}\in \mathbb{R}).
$$
The last equality implies that the sequence $Z_{[1]},Z_{[2]},\ldots$ forms a Markov chain. The Markov property also holds in the continuous case.

The following limit
\begin{equation}\label{3.1}
\lim_{x\rightarrow
r_H}\frac{G(y)-F(x,y)}{1-H(x)}=\beta(y)\in[0,1]
\end{equation}
is considered in Bairamov and Stepanov
(2010), (2011). They showed that if $\beta(y)=0$ for any $y<r_G=\sup\{y\in \mathbb{R}:\:G(y)<1\}$, then
$$
Y_{[n-k,n]}\stackrel{p}{\rightarrow}r_G\quad (n\rightarrow\infty).
$$
With some additional conditions on $F$ one can formulate a strong
limit law for $Y_{[n,n]}$.

\begin{theorem}\label{theorem3.1} The convergence
$$
Y_{[n,n]}\stackrel{a.s.}{\rightarrow}r_G\quad
(n\rightarrow\infty)
$$
holds true iff
\begin{equation}\label{3.2}
\int_{\mathbb{R}}\frac{G(y)-F(x,y)} {(1-H(x))^2}\ [dH(x)-F(dx,y)]<
\infty
\end{equation}
for any $y<r_G$. 
\end{theorem}
\begin{gproof}{} 
By symmetry and independence, we have
$$
P(A_n)=P(Y_{[n,n]}\leq y)=
$$$$
n\int_\mathbb{R}\int_{l_G}^yP(X_2\leq x,\ldots,X_{n}\leq x)
F(dx,dv).
$$
It follows that
\begin{equation}\label{3.3}
P(Y_{[n,n]}\leq y)=n\int_\mathbb{R}H^{n-1}(x)
F(dx,y).
\end{equation}

By the argument that is used for obtaining (\ref{3.3}), one can get
$$
P(A_n^cA_{n+1})=P(Y_{[n,n]}>y,Y_{[n+1,n+1]}\leq y)=
$$
$$
n\int_{\mathbb{R}}H^{n-1}(x)(G(y)-F(x,y))[dH(x)-F(dx,y)].
$$
Then
$$
\sum_{n=1}^\infty P(A_n^cA_{n+1})=
$$
$$
\int_{\mathbb{R}}\frac{G(y)-F(x,y)} {(1-H(x))^2}\
[dH(x)-F(dx,y)].
$$
Theorem~\ref{theorem3.1} readily follows from Lemma~\ref{lemma2.2}.
\end{gproof}
\begin{remark}\label{remark3.1} Observe that under condition
(\ref{3.2}) we have $\beta(y)=0$.
\end{remark}

\subsection{Strong Limit Results  for the $F^\alpha$-Scheme}
Let in the following, $X_1, X_2, \ldots, X_n$ be independent
continuous random variables with distribution functions
$F^{\alpha_1}, F^{\alpha_2},\ldots, F^{\alpha_n}$, where
$\alpha_i>0\ (i\geq2)$ and $\alpha_1=1$. These settings are known
as the $F^\alpha$-scheme, in which the independent variables $X_1,
X_2, \ldots, X_n$ have the same support.  When
$\alpha_i=1$, the $F^\alpha$-scheme reduces to the case when the
variables $X_i$ are independent and identically distributed.
The $F^\alpha$-scheme was first considered by  Yang (1975) as a model of a non-stationary sequence of independent random variables. Later it was discussed by Nevzorov (1985, 1986), Pfeifer (1989, 1991), Bairamov and Stepanov (2013), and others.

{\bf Strong Limit Results for  Maxima} Let in the following, $M_n=\max\{X_1,\ldots,X_n\}$, $A_n=1+\alpha_2+\ldots+\alpha_n$ and $x_n$ be a nondecreasing sequence of real numbers. It is easily found  that $P(M_n\leq x_n)=F^{A_n}(x_n)$. It follows from the Borel-Cantelli lemma that if
\begin{equation}\label{4.1}
\sum_{n=1}^\infty F^{A_n}(x_n)<\infty,
\end{equation}
then $P(M_n\leq x_n\ i.o.)=0$. One can show that the sequence $M_n\ (n\geq 1)$ forms a Markov chain. Lemma~\ref{lemma2.3} allows us to formulate the following result.
\begin{proposition}\label{proposition4.1}
The equality
$$
P(M_n\leq x_n\ i.o.)=\left\{
\begin{array}{cc}
0\\
1 
\end{array}
\right.
$$
holds true iff 
\begin{equation}\label{4.2}
\sum_{n=1}^\infty F^{A_n}(x_n)[1-F^{\alpha_{n+1}}(x_{n+1})]\left\{
\begin{array}{cc}
<\infty\\
=\infty.
\end{array}
\right.
\end{equation}
\end{proposition}
Observe that if $F^{\alpha_n}(x_n)\rightarrow 1$, then the series in (\ref{4.2}) converges under weaker conditions than the serious in (\ref{4.1}). One can propose  examples when the Borel-Cantelli lemma fails to produce strong limit results and Lemma~\ref{lemma2.3}  produces  strong limit results in forms of "iff`` statements. One of such examples is given below.

\begin{example}\label{example4.1}
Let $F$ be the unit uniform distribution. Let us choose 
$$
x_n=1-\frac{\log\log n}{n},\quad  \alpha _n=\gamma (1+1/n)\quad \mbox{and}\quad A_n\approx \gamma(n+\log n).
$$
where $\gamma>0$. Observe that $ F^{A_n}(x_n)\rightarrow 0$ and $\sum_{n=1}^\infty  F^{A_n}(x_n)=\infty$ for any    $\gamma>0$.
It follows that the Borel-Cantelli lemma can not help us in this case. 

The series $\sum_{n=1}^\infty F^{A_n}(x_n)[1-F^{\alpha_{n+1}}(x_{n+1})]$ behaves like the series  $\sum_{n=1}^\infty\frac{\log \log n}{n(\log n)^\gamma }$ when $\gamma>0$.  Lemma~\ref{lemma2.3} then states that 
$$
P(M_n\leq 1-\frac{\log\log n}{n}\ \ i.o.)=\left\{
\begin{array}{cc}
0\\
1 
\end{array}
\right.
$$
iff 
$$\gamma\in\left\{
\begin{array}{cc}
(1,\infty)\\
(0,1].
\end{array}
\right.
$$
\end{example}

{\bf Could the newcomer be a maximum?} Observe that if $X\sim F,\ Y\sim F^\alpha$ and $\alpha>1\
(0<\alpha<1)$, then $Y\geq_{st}X\ (Y\leq_{st}X)$, where $st$ means
the stochastic comparison of $X$ and $Y$. That way, choosing
$\alpha_n$ as an increasing sequence for the $F^\alpha$-scheme, we
shift  probability masses $F^{\alpha_n}$ towards the right end of
the common support. This gives new members of the sequence
$X_1,\ldots,X_n,X_{n+1}$ (we address to $X_{n+1}$ as to a new
member) better chances to become  maxima. We will show  that  for properly chosen  $\alpha_n$ and all large enough
$n$ each consecutive sample observation can be a maximum. On the
contrary, choosing the sequence $\alpha_n$ as decreasing we can
"prohibit" newcomers to be maxima.

Suppose that $\alpha_n\rightarrow\infty$, which means that
$A_n\rightarrow\infty$. As was pointed out, under
the proper rate of increase of $\alpha_n$, each new sample
observation can be a maximum. The corresponding conditions are
proposed below in Proposition~\ref{proposition5.1} and Remark~\ref{remark3.2}.
\begin{proposition}\label{proposition5.1}
Let $\alpha_n\rightarrow\infty$ and $A_n/A_{n+1}\rightarrow 0$. Then
$$
P(M_n-X_n>0\ i.o.)=\left\{
\begin{array}{cc}
0\\
1 
\end{array}
\right.
$$
iff 
\begin{equation}\label{5.1}
\sum_{n=1}^\infty A_n/A_{n+1}\left\{
\begin{array}{cc}
<\infty\\
=\infty.
\end{array}
\right.
\end{equation}
\end{proposition}

\begin{gproof}{}
Let us denote $B_n=\{M_n-X_n>0\}$ and  $B^c_n=\{M_n=X_n\}$. Then
\begin{equation}\label{5.2}
P(B_n)=\int_{\mathbb{R}}
F^{\alpha_n}(x)dF^{A_{n-1}}(x)=A_{n-1}/A_n
\end{equation}
and
$$
P(B_nB_{n+1}^c)=\int_{\mathbb{R}}
F^{\alpha _n}(x)(1-F^{\alpha_{n+1}}(x))dF^{A_{n-1}}(x)=
$$
\begin{equation}\label{5.3}
\frac{\alpha_{n+1}}{A_{n+1}}\cdot \frac{A_{n-1}}{A_n}\sim \frac{A_{n-1}}{A_n}.
\end{equation}
The result follows from Lemma~\ref{lemma2.3},  (\ref{5.2}) and (\ref{5.3}).
\end{gproof}
\begin{remark}\label{remark3.2}
If the series in (\ref{5.1}) is convergent, then $M_n-X_n\rightarrow0\ a.s.$, i.e. for all large enough $n$  with probability one each new sample member is a maximum. This, in particular, happens when $\alpha_n=n^{2n}$.
\end{remark}
 
As was mentioned above, a properly chosen $\alpha_n$ can
"prohibit" newcomers to be maxima. This issue is discussed in
the rest of our work.
\begin{proposition}\label{proposition5.2}
\begin{enumerate}
Let
\begin{equation}\label{5.4}
\alpha_n/A_n\rightarrow0.
\end{equation}
Then
$$
P(M_n=X_n\ i.o.)=\left\{
\begin{array}{cc}
0\\
1 
\end{array}
\right.
$$
iff 
\begin{equation}\label{5.5}
\sum_{n=1}^\infty \alpha _n/A_{n}\ \left\{
\begin{array}{cc}
<\infty\\
=\infty.
\end{array}
\right.
\end{equation}
\end{enumerate}
\end{proposition}
\begin{gproof}{}
The proof is similar to the proof of Proposition~\ref{proposition5.1}. One should  take $C_n=B^c_n, C_n^c=B_n$ and apply Lemma~\ref{lemma2.3} to the sequence of events  $C_n$.
\end{gproof}
\begin{remark}\label{remark3.3}
If the series in (\ref{5.5}) is convergent, then $M_n>X_n\ a.s.$, i.e. for all large enough $n$ with probability one no new sample member  can be a maximum. This, in particular, happens when $\alpha_n=n^{-2}$.
\end{remark}

\section*{References} 
\begin{description}

\item Bairamov, I. and Stepanov, A. (2010).\ Numbers of
near-maxima for the bivariate case,  {\it Statistics  Probability
Letters}, {\bf 80} (3), 196--205.

\item Bairamov, I. and Stepanov, A. (2011).\ Numbers of near
bivariate  record-concomitant observations spacings,  {\it Journal
of Multivariate Analysis},  {\bf 102} (5), 908--917.

\item  Bairamov, I. and Stepanov, A. (2013).\ Numbers of near-maxima
for $F^\alpha$-scheme, Statistics, {\bf 47}, 191--201.

\item Balakrishnan, N., Stepanov, A. (2010).\ Generalization of
Borel-Cantelli lemma - {\it The Mathematical Scientist}, {\bf 35}
(1), 61--62.

\item Barndorff-Nielsen, O. (1961).\ On the rate of growth of the
partial maxima of a sequence of independent identically
distributed random variables. {\it Math. Scand.},  {\bf 9},
383--394.

\item Bhattacharyya, B. B. (1974).\ Convergence of sample paths of
normalized sums of induced order statistics, {\it Ann. Statist.},
{\bf 2}, 1034-1039.

\item Chung, K.L. and Erd\"os, P. (1952).\ On the application of the
Borel-Cantelli lemma. {\it Trans.  Amer. Math. Soc.},  {\bf 72},
179--186.

\item Chandra, T.K. (2012).\ {\it The Borel-Cantelli Lemma}, Springer Briefs in Statistics.

\item David, H. A. (1973).\ Concomitants of order statistics, {\it
Bull. Inst. Internat. Statist.}, {\bf 45} (1), 295--300.

\item David, H. A. and Galambos, J. (1974).\ The asymptotic theory
of concomitants of order statistics, {\it J. Appl. Prob.}, {\bf
11}, 762--770.

\item David, H.A., Nagaraja, H.N. (2003).\ {\it Order Statistics}. Third
edition,  John Wiley \& Sons, New York.

\item Egorov, V. A. and Nevzorov, V. B. (1984).\ Rate of
Convergence to the Normal Law of Sums of Induced Order Statistics,
{\it Journal of Soviet Mathematics} (New York), {\bf 25}, 1139–-
1146.

\item Erd\"os, P. and R\'enyi, A. (1959).\ On Cantor's series with
convergent $\sum 1/q_n$. {\it Ann. Univ. Sci. Budapest. Sect.
Math.}, {\bf 2}, 93--109.

\item Frolov, A.N. (2012).\ Bounds for probabilities of unions of events and the Borel-Cantelli lemma. {\it Statist. Probab. Lett.}, {\bf 82}, 2189--2197.

\item Goel, P.K. and Hall, P. (1994).\ On the average difference between concomitants and order statistics,
{\it The Annals of Probability}, {\bf 22} (1),
 126--144.
 
\item Kochen, S.B. and Stone, C.J. (1964).\ A note on the
Borel-Cantelli lemma. {\it Illinois J. Math.},  {\bf 8}, 248--251.

\item Kounias, E.G. (1968).\ Bounds for the probability of a union, with applications. {\it Ann. Math. Statist.}, {\bf 39}, 2154--2158.

\item Lamperti, J. (1963).\  Wiener's test and Markov chains. {\it J.Math.Anal.Appl.}, {\bf 6}, 58–-66.

\item L\'evy,  P. (1937).\ {\it Theorie de l' addition des variables aleatoires}, Gauthier-Villars, Paris.

\item Martikainen, A.I., Petrov, V.V., (1990).\ On the
Borel–Cantelli lemma. {\it Zapiski Nauch. Semin. Leningrad. Otd.
Steklov Mat. Inst.}, {\bf 184}, 200–-207 (in Russian). English
translation in:  (1994). {\it J. Math. Sci.}, {\bf 63}, 540–-544.

\item M\'ori T.F. and Sz\'ekely, G.J. (1983).\  On the Erdos-Renyi generalization of the Borel-Cantelli lemma. {\it Studia Sci. Math. Hunger.}, {\bf 18}, 173--182.

\item Nagaraja, H. N. and David, H. A. (1994).\ Distribution of
the maximum of concomitants of selected ordered statistics, {\it
Ann. Statist.}, {\bf 22}, 478--494.

\item Nevzorov,  V.B. (1985).\  On record times and inter-record times for sequences of nonidentically distributed random variables,
{\it Zap. Nauch. Sem. POMI(LOMI)}, {\bf 142}, 109-–118.

\item Nevzorov, V.B. (1986).\ Two characterizations using records, {\it Lecture Notes in Mathematics}, {\bf 1233}, Springer, Berlin,
79–-85.

\item Petrov, V.V. (2002).\ A note on the Borel-Cantelli lemma.
{\it Statist. Probab. Lett.}, {\bf 58}, 283--286.

\item Petrov, V.V. (2004).\ A generalization of the Borel-Cantelli
Lemma. {\it Statist. Probab. Lett.}, {\bf 67}, 233--239.

\item Pfeifer, D. (1989).\ Extremal processes, secretary problems and the 1/e law, {\it J. Appl. Probab.}, {\bf 27}, 722–733.

\item Pfeifer, D. (1991).\ Some remarks on Nevzorov's record model, {\it Adv. Appl. Probab.}, {\bf 23}, 823--834.

\item Sen, P.K. (1981).\ Some invariance principles for mixed rank statistics and induced order statistics and some application, {\it Communications in Statistics - Theory and Methods},
{\bf 10} (17).

\item Spitzer, F. (1964).\ {\it Principles of Random Walk}, Van Nostrand, Princeton.

\item Yang, M.C.K. (1975).\ On the distribution of the inter-record times in an increasing population, {\it J. Appi. Probab.}, {\bf 12},
148–-154.

\end{description}
\end{document}